\documentclass[12pt]{article}
\usepackage{graphics,epsfig}

\title{The role of inflexible minorities in the  breaking of democratic opinion dynamics}

\author{Serge Galam$^1$\thanks{E-mail: serge.galam@polytechnique.edu} and Frans Jacobs$^2$\thanks{E-mail: f.j.a.jacobs@biology.leidenuniv.nl}\\
$^1$Centre de Recherche en \'Epist\'emologie Appliqu\'ee (CREA),\\
\'Ecole Polytechnique, 1 Rue Descartes, 75005 Paris, France\\
$^2$Section Theoretical Biology, Institute of Biology, Leiden University\\ Kaiserstraat 63, NL-2311 GP Leiden, The Netherlands }

\date{February 20, 2007}

\begin{document}
\maketitle

\begin{abstract}
We study the effect of inflexible agents on two state opinion dynamics. The model operates via repeated  local updates of random grouping of agents. While floater agents do eventually flip their opinion to follow the local majority, inflexible agents keep  their opinion always unchanged. It is a quenched individual opinion. In the bare model  (no inflexibles), a separator at $50\%$  drives the dynamics towards either one of two pure attractors, each associated with a full polarization along one of the opinions. The initial majority wins. The existence of inflexibles for only one of the two opinions
is found to shift the separator at a lower value than $50\%$ in favor of that side. Moreover it creates an incompressible minority around the inflexibles, one of the pure attractors becoming a mixed phase attractor. In addition above a threshold of $17\%$ inflexibles make their side sure of winning whatever the initial conditions are. The inflexible minority wins. An equal presence of inflexibles on both sides restores the balanced dynamics with again a separator at $50\%$ and now two mixed phase attractors on each side. Nevertheless, beyond  $25\%$ the dynamics is reversed with a unique attractor at a fifty-fifty stable equilibrium. But a very small advantage in inflexibles results in a decisive lowering of the separator at the advantage of the corresponding opinion. A few percent advantage does guarantee to become majority with one single attractor. The model is solved exhaustedly for groups of size 3. 
\end{abstract}

PACS's numbers: 02.50.Ey, 05.45.-a, 9.65.-s, 87.23.Ge \\\\

Opinion dynamics has become a very active subject of research \cite{sznajd, sorin, red-1, red-2, slanina, neigbhor-1, neigbhor-2, ausloos, koree} in sociophysics \cite{strike,socio}. Most works consider two state models which lead to the disappearance of one of the two opinions. They use local updates in odd size groups which result in the initial majority victory. A unifying frame was shown to include most of these models  \cite{unify}. Continuous extensions \cite{deffuant, krausse} and three state models \cite{3-choices} have been also investigated.

However,  including an inertia effect in even size local updates groups, the initial minority may win the competition spreading over the entire population.  The inertia effect means that in an update even size group at a tie, the opinion which preserves the Status Quo is selected locally by all the group members \cite{mino, hetero}. When an opinion represents a vote intention, the model allows to make successful prediction in real voting cases like for the 2005 french referendum \cite{lehir}. 

At contrast it is found that including contrarian behavior leads to the reversal of the dynamics with a stable equilibrium at exactly fifty-fifty whatever the initial conditions are. A contrarian is an agent who makes up its opinion by choosing the one minority opinion, either the local minority within its update group \cite{contra-1} or the global minority according to polls \cite{contra-2}. It was used to explain and predict the occurrence of a recent series of hung elections in democratic countries  \cite{contra-1}.

In addition to contrarian behavior \cite{contra-1, contra-2, contra-3, contra-4}, another type of behavior is also quite current while dealing with real opinion dynamics, it is the inflexible attitude. At contrast to floater agents who do eventually flip their opinion to follow the local majority, inflexible agents 
keep their opinion always unchanged. The inflexible attitude is a quenched individual state. Surprisingly,  it has not been studied so far. It is the subject of this article to investigate the inflexible effect on the associate opinion dynamics. To confront our results to any real situation requires to have an estimate of the various densities of inflexibles, which could be extracted in principle from appropriate polls.

\begin{figure}
\centering
\includegraphics[width=.5\textwidth]{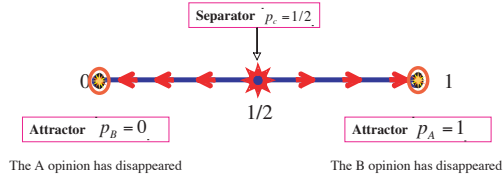}
\caption{The bare model with only floaters. The initial majority is conserved and increased to eventually invade the whole population.}
\label{bare-1}
\end{figure}

In the bare model, where no inflexible  is present, denoting A and B the two competing opinions and $p_t$ the density of A at time $t$, the flow diagram of the dynamics is monitored by a separator at $p_c=50\%$. it drives the dynamics towards either one of two pure attractors, $p_B=0$ where the A opinion has totally disappeared, and $p_A=1$ where the A opinion has totally invaded the whole population. It is shown in  Fig. (\ref{bare-1}). The initial majority always wins. 

\begin{figure}
\centering
\includegraphics[width=.5\textwidth]{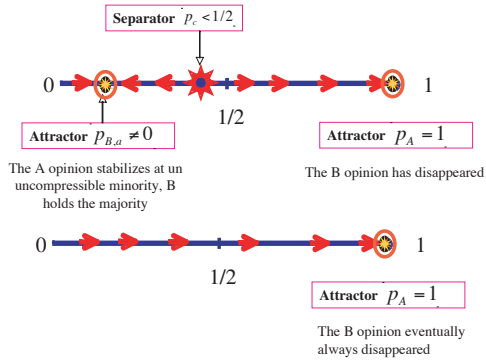}
\caption{One side inflexible at low density. In the upper part inflexibles shift the separator to a lower value than $50\%$ at the advantage of their side. Moreover, the associated opinion never disappears but at minimum stabilizes at some stable minority value $p_{B,a}$. The associated opinion can now invade the whole population even when it starts  at an initial value lower than $50\%$ within some appropriate range. The lower part shows that beyond  $17\%$ in the density of inflexibles, the separator and the mixed phase attractor have vansihed after they have coalesced. At any initial condition, the A wins and eventually invades the whole population.}
\label{inf-1}
\end{figure}

The existence of inflexibles for only one of the two opinions, for instance opinion A, is found to shift the separator at a lower value than $50\%$ in favor of that side. Moreover it creates an incompressible minority around the inflexibles, one of the pure attractors, here $p_B$, becoming a mixed phase attractor, where opinion B holds the majority but with a stable A minority, $p_B=0\rightarrow p_{B,a}\neq 0$. See the upper part of Fig. (\ref{inf-1}). In addition, increasing the one side inflexible density above some threshold ($17\%$ for update group of size 3) inflexibles make the separator and the mixed phase attractor to coalesce and thus cancel each other to both disappear. Their side becomes certain of winning whatever the initial conditions are. The inflexible minority wins as illustrated in the lower part of Fig. (\ref{inf-1}). 

\begin{figure}
\centering
\includegraphics[width=.5\textwidth]{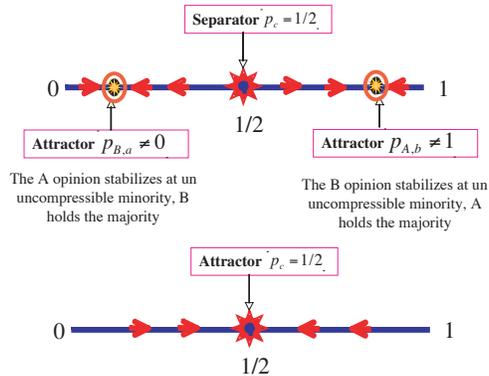}
\caption{Equal presence of inflexibles on both sides. In the upper part the balanced dynamics is restored with the separator back at $p_c=50\%$.  Now two mixed phase attractors $p_{B,a}\neq 0$ and $p_{A,b}\neq 1$  are located on each side of the separator. Nevertheless, in the lower part, beyond  $25\%$ they both coalesce with the separator, which at once becomes the  unique attractor. The dynamics is reversed with a coexistence of both opinions at a fifty-fifty stable equilibrium.}
\label{inf-2}
\end{figure}

However an equal presence of inflexibles on both sides is shown to restore the balanced dynamics with again the separator at $p_c=50\%$ and now two mixed phase attractors  $p_{B,a}\neq 0$ and $p_{A,b}\neq 1$ on each side as seen in the upper part of Fig. (\ref{inf-2}). Nevertheless, beyond  $25\%$ the dynamics is reversed with a unique attractor at a fifty-fifty stable equilibrium. See the lower part of Fig. (\ref{inf-2}).  

\begin{figure}[t]
\centering
\includegraphics[width=.5\textwidth]{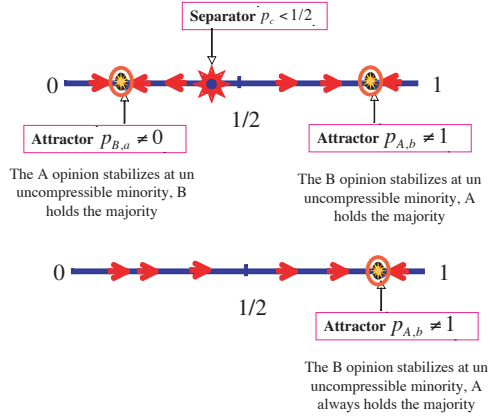}
\caption{Unequal densities of inflexibles. The upper part shows a rather small difference in inflexibles, which  results in a decisive lowering of the separator at the advantage of the corresponding larger side. The lower part shows the case of a few percent advantage, which does grant the victory.}
\label{inf-3}
\end{figure}

But again, a very small advantage in inflexibles results in a decisive lowering of the separator at the advantage of the corresponding opinion as shown in the upper part of  Fig. (\ref{inf-3}). In addition the lower part of Fig. (\ref{inf-3}) shows that a few percent advantage does grant the victory. 

We now solve analytically the problem for local update groups of size 3. Initial proportions at time $t$ of both opinion are respectively $p_t$ and  $(1-p_t)$ where each agent does have an opinion. On the A side, at any time the associated agents holder are divided among a fixed and constant proportion of inflexibles $a$, they always keep on opinion A, and a varying density of floaters $p_t-a$. The floaters do shift opinion depending on their local update group composition. Similarly, on the opposite side B, the agent holder contains a fixed and constant proportion of inflexibles $b$ with a density of 
 $(1-p_t-b)$ floaters. 
 
Dealing with densities we have the constraints $0\leq a \leq 1$, $0\leq b \leq 1$,  $0\leq a+b \leq 1$ and $a\leq p_t \leq 1-b$. To make the notations more practical we introduce the difference in inflexible densities $x$ to write $a\equiv b+x$ with $-b\leq x \leq 1-2b$. The value of $x$ may be negative to account for an advantage to the B opinion. A positive value corresponds to an advantage to A. The two external parameters of the problem are thus $b$ and $x$.

Then at time $t$ people are grouped randomly by three and a local majority rule is applied separately within each local group. At time $t+1$ within each group all floaters who held the minority opinion do shift to the local majority one. However inflexibles do not shift their opinion. Dealing with three agents, the only subtle cases are the ones where 2 agents sharing the same opinion are  against the third who holds the other one. In case it is a floater the minority agent joins the majority, otherwise being an inflexible, it does shift opinion and keeps the minority opinion. A detailed counting of all cases leads to write at time $t+1$ for the new proportion of opinion A,
\begin{equation}
p_{t+1}=p_t^3+3p_t^2\left ((1-p_t-b)+\frac{2}{3} b \right )+3(1-p_t)^2\left ( \frac{1}{3} a \right ),
\end{equation}
which simplifies to
\begin{equation}
p_{t+1}=-2 p_t^3+p_t^2 (3+x)-2(b+x) p_t +b+x.
\label{b-x}
\end{equation}

After one update, all agents are reshuffled before undergoing a second redistribution among new random groups of three agents each. Now $p_{t+1}$ plays the role of $p_t$ before, and a new density $p_{t+2}$ is obtained.  The process is repeated some number $n$ of times leading to the density $p_{t+n}$ of agents sharing opinion A and $1-p_{t+n}$ of agents sharing opinion B. It is worth to stress that the respective proportions of inflexibles $a$ and $b$ are unchanged and independent of the value of $n$.

While the reshuffling frame has been viewed as belonging to a mean field treatment \cite{slanina, neigbhor-1, neigbhor-2}, it has demonstrated to indeed create a new universality class \cite{reshuffle}.

Before proceeding we review the bare model, i.e., no inflexible is present ($a=b=0$) and all agents are floaters. From Eq. (\ref{b-x}) one cycle of local opinion updates via three persons grouping leads to the new distribution of vote intention as,
\begin{equation}
p_{t+1}=p_t^3+3p_t^2(1-p_t) ,
\label{0-0}
\end{equation}
whose dynamics is monitored by the unstable fixed point separator located at $p_c=\frac{1}{2}$. It  separates the respective basins of attraction of the two pure phase stable point attractors at $p_A=1$ and $p_B=0$.  Accordingly  $p_{t+1}>p_t$  if $p_{t+1}>\frac{1}{2}$ and $p_{t+1}<p_t$ 
if $p_{t+1}<\frac{1}{2}$ as shown in Fig. (\ref{bare-1}). The initial majority wins.

For instance starting at $p_t=0.45$  leads successively after 5 updates to the series 
$p_{t+1}=0.43, p_{t+2}=0.39, p_{t+3}=0.34, p_{t+4}=0.26, p_{t+5}=0.17$
with a continuous 
decline in A support. Adding 3 more cycles would result in zero A support with $p_{t+6}=0.08, p_{t+7}=0.02$ and $p_{t+8}=0.00$. Given any initial distribution of opinions, the random local 
opinion update leads toward a total polarization of the collective opinion. 
Individual and collective opinions stabilize simultaneously along the same and 
unique vote intention either A or B. 

The update cycle number to reach either one of the two stable attractors can be 
evaluated from Eq. (\ref{b-x}). It depends on the distance of the initial densities from 
the unstable point attractor. 
However, every update cycle takes some time length, which may correspond in real terms
to some number of days. 
Therefore, in practical terms the required time to eventually complete the 
polarization process is much larger than any public debate  duration, thus preventing 
it to occur. Accordingly, associate elections never take place at the stable 
attractors. From the above example at $p_t=0.45$, two cycles yield a result of 
$39\%$ in favor of A and $61\%$ in favor of B. One additional update cycle makes 
$34\%$ in favor of A and $66\%$ in favor of B.

We can now insert the existence of inflexibles. To grasp fully its social meaning we will introduce it in several steps. For the first one, inflexibles are present only on one side, say A. We thus have $b=0$ which yields $a=x$. Eq. (\ref{b-x}) becomes
\begin{equation}
p_{t+1}=-2 p_t^3+p_t^2 (3+x)-2x p_t +x.
\label{0-x}
\end{equation}

Solving the associated fixed point Equation $p_{t+1}=p_t$ yields the three solutions 
\begin{equation}
p_{B,a}=\frac{1}{4}\left (1+x-\sqrt {1-6x+x^2} \right) ,
\label{x-p1}
\end{equation}

\begin{equation}
p_{c}=\frac{1}{4}\left (1+x+\sqrt {1-6x+x^2} \right) ,
\label{x-p2}
\end{equation}
and $p_A=1$ to be compared to the bare results ($x=0$) $p_B=0$,  $p_{c}=\frac{1}{2}$ and $p_A=1$.
While $p_B$, and $p_{c}$ have been shifted toward one another, $p_A$ stayed unchanged as in the upper part of Fig. (\ref{inf-1}). 

\begin{figure}[t]
\centering
\includegraphics[width=.5\textwidth]{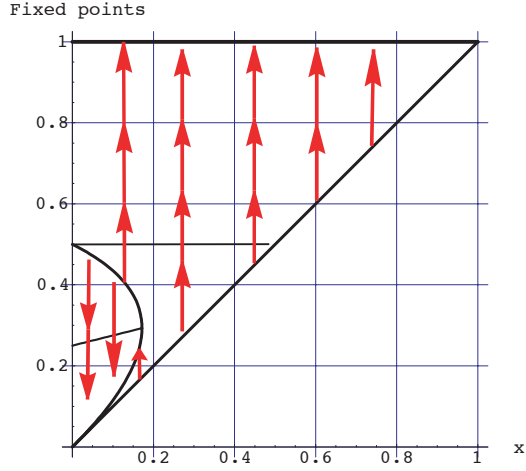}
\caption{One sided inflexibles fixed points as a function of their density $x$.  One line of attractors $p_A=1$. In the regime $x<0.17$ the left upper part of the curved line is a line of separator (Eq. (\ref{x-p1})) while the lower part is a line of attractor (Eq. (\ref{x-p2})). Both are symmetrical with respect tot the line $\frac{1+x}{4}$ at which they eventually coalesce at $x_c=3-2\sqrt {2}\approx 0.17$. The diagonal line delimits the floater region for A holders since $p\leq x$.
As soon as  $x>0.17$ the victory is granted for opinion A.}
\label{main-1}
\end{figure}

From above expressions an increase in $x$ gets closer the attractor $p_{B,a}$ and the separator $p_{c}$ before they coalesce at $x_c=3-2\sqrt {2}\approx 0.17$, and there disappear as seen in Fig. (\ref{main-1}). The attractor $p_A$ stays independent of $x$. Therefore for $x>0.17$ the unique fixed point of the dynamics is the attractor $p_A=1$. Any initial support in A leads to its victory.

\begin{figure}
\centering
\includegraphics[width=.5\textwidth]{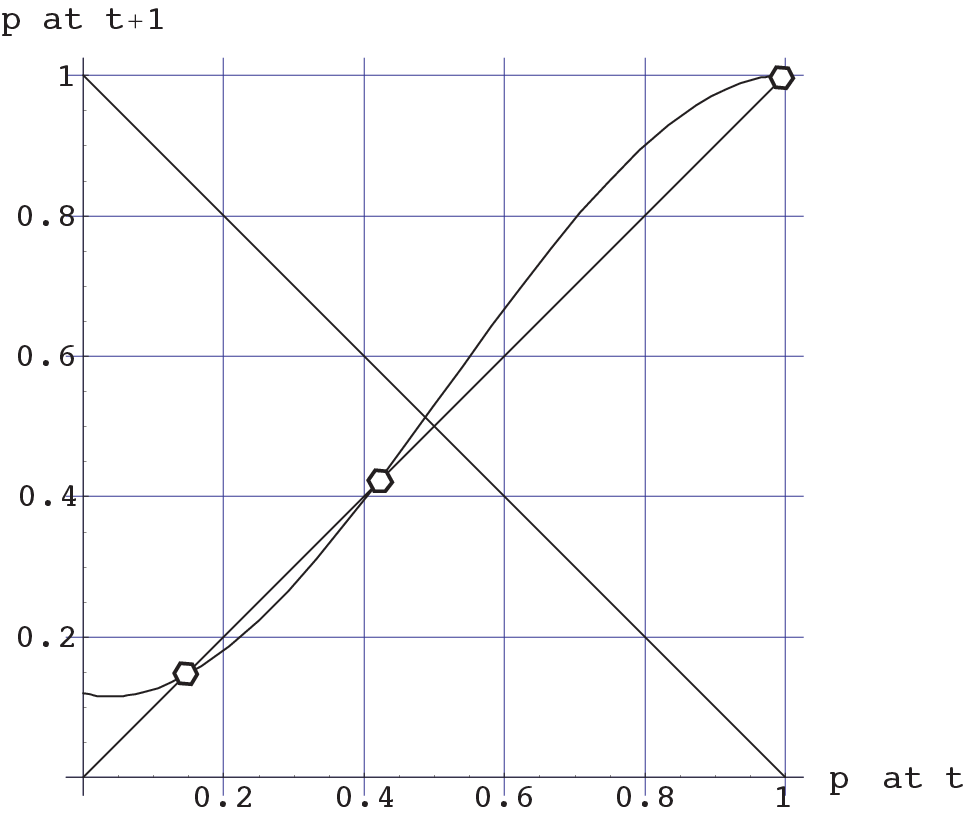}\hfill
\includegraphics[width=.5\textwidth]{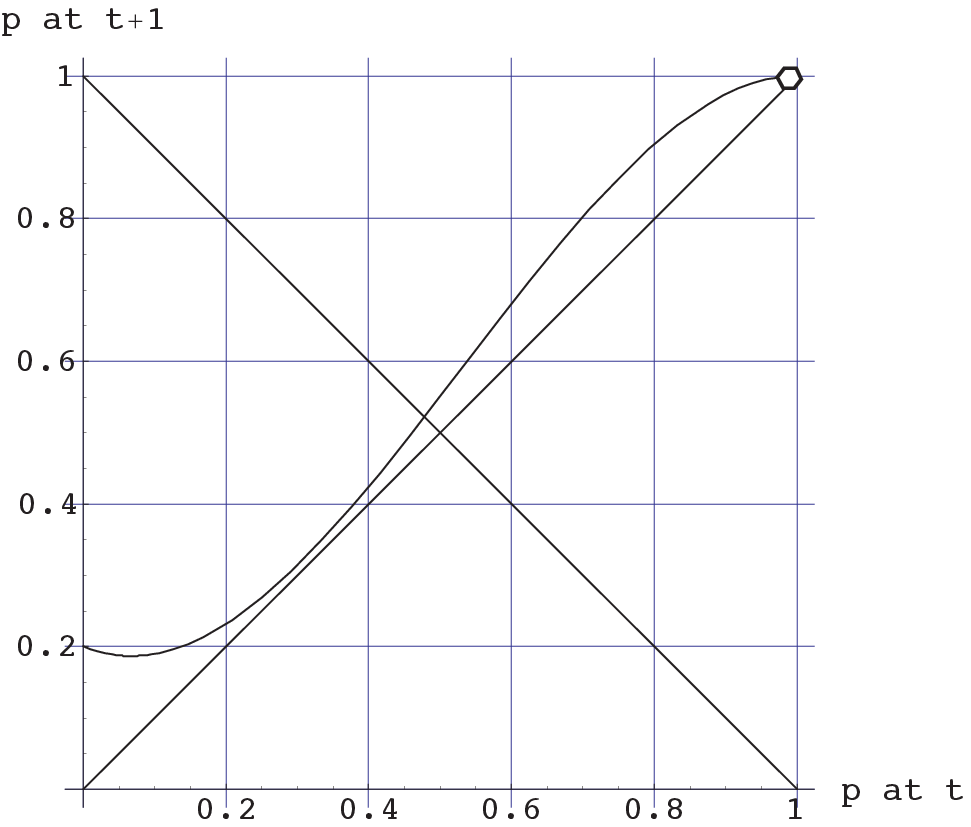}
\caption{One sided inflexibles. The left  part corresponds to $x<0.17$ of inflexibles in favor of opinion A. The right part shows the case of $x>0.17$ which does grant the victory. to opinion A.}
\label{main-2}
\end{figure}

Fig. (\ref{main-2}) shows the variation of $p_{t+1}$ as a function of $p_t$ for these two regimes. It is worth to note that in the second regime the dynamics of the winning inflexible minority is slowed down in some window of support before it starts to increase at a speedy path.

For instance $p_t=0.20$  leads successively  to the series 
$p_{t+1}=0.23, p_{t+2}=0.25, p_{t+3}=0.27, p_{t+4}=0.29, p_{t+5}=0.30, p_{t+6}=0.32,
p_{t+7}=0.33, p_{t+8}=0.34, p_{t+9}=0.36, p_{t+10}=0.38, p_{t+11}=0.40,
p_{t+12}=0.42, p_{t+13}=0.45, p_{t+14}=0.49, p_{t+15}=0.53, p_{t+16}=0.59,
p_{t+17}=0.67, p_{t+18}=0.77, p_{t+19}=0.87, p_{t+20}=0.96, p_{t+21}=1.00$,
with a continuous increase  in A support. However $15$ updates are necessary for A  to reach the majority from its initial $20\%$. Before, at $x=0$, $8$ updates were reducing a $45\%$ support to zero while now $15$ are required to gain $30\%$. 

\begin{figure}[t]
\centering
\includegraphics[width=.7\textwidth]{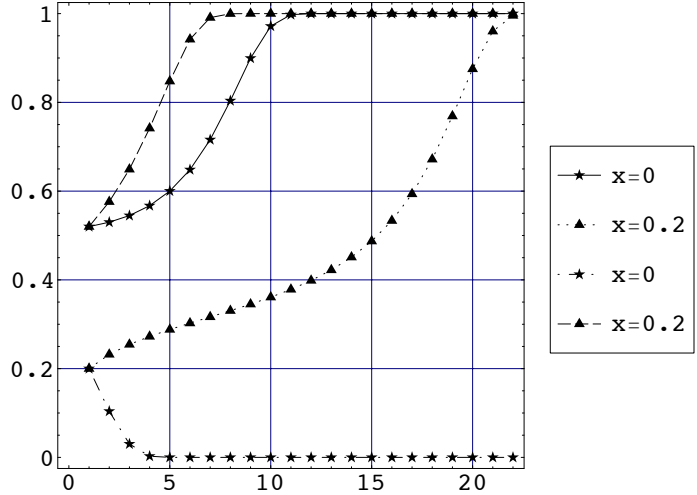}
\caption{Comparison of the update series from two initial supports $p_t=0.52$ and $p_t=0.48$ for the pure floater case $x=0$ and one sided inflexible with a density  $x=0.20$ above the threshold $x_c\approx 0.17$. In the latter case the victory is granted for opinion A although it starts from such a lower support  of $20\%$. Nevertheless the process is rather slow.}
\label{main-3}
\end{figure}

In terms of real time durations, a number of $15$ updates may imply many months. 
Fig. (\ref{main-3}) shows two initial supports  $p_t=0.20$ and $p_t=0.52$ for respectively $x=0$ and $x=0.20$. The differences in the associated dynamics are drastic.

We note that setting $x=-b$ defines the symmetric situation with inflexibles only on side B. We then have $a=0$ and $b$ for the respective densities of inflexibles. Above results then apply to the B opinion with the variable $b$ playing the role of $x$.

At this point to have inflexibles on its side appears to be a decisive step towards leading the opinion competition. Accordingly both opinions are expected to have inflexibles. in case of a symmetric presence of inflexibles on both sides with $x=0$ and $b\neq 0$, i.e., $a=b\neq 0$. In addition, since the total density of both side inflexibles is $2b$, the variable $b$ must obeys $b\leq \frac{1}{2}$. Eq. (\ref{b-x}) becomes
\begin{equation}
p_{t+1}=-2 p_t^3+3p_t^2-2 b  p_t +b,
\label{b-0}
\end{equation}
whose fixed points are 
\begin{equation}
p_{B,a}=\frac{1}{2}\left (1-\sqrt {1-4b} \right) ,
\label{b-p1}
\end{equation}

\begin{equation}
p_{A,b}=\frac{1}{2}\left (1+\sqrt {1-4b} \right) ,
\label{b-p2}
\end{equation}
and $p_c=\frac{1}{2}$. The symmetry restoring has put back the separator at $\frac{1}{2}$ independently of $b$. The two mixed phase attractors $p_{B,a}$ and $p_{B,a}$ are now symmetric and move towards $p_c$ as a function of increasing $b$. It is again the initial majority which wins the competition.

Nevertheless at $a=b=\frac{1}{4}$ the dynamics is turned up side down with $p_{B,a}$ and $p_{A,b}$ merging at $p_c=\frac{1}{2}$, which at once becomes an attractor and the unique fixed point of the dynamics. Any initial condition leads to a hung equilibrium with an identical support of $50\%$ for both opinions. 

\begin{figure}
\centering
\includegraphics[width=.5\textwidth]{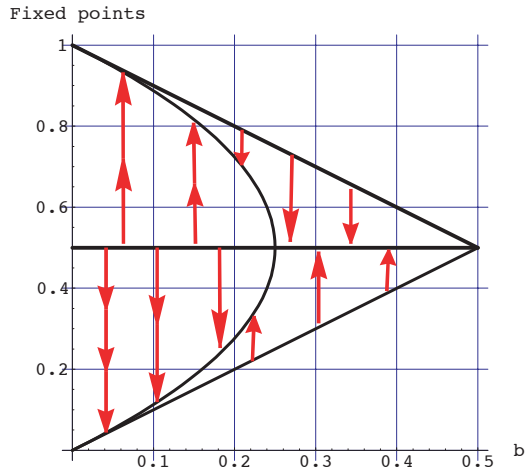}
\caption{Two side symmetric inflexibles fixed points as a function of their density $b$.  The first part of the line  $p_c=\frac{1}{2}$ till $b=\frac{1}{4}$ is a separator. From there, it becomes the unique attractor of the dynamics. The left curved line is a line of mixed phase attractors $p_{A,b}$ (upper part, Eq. (\ref{x-p1})) and $p_{B,a}$ (lower part, Eq. (\ref{x-p2})). Both are symmetrical with respect tot the line $\frac{1}{2}$ at which they eventually coalesce at $b_c=\frac{1}{4}$. The two lines $b$ and $(1-b)$ delimits the floater region for A holders since $p\geq b$ with $b\leq \frac{1}{2}$. 
As soon as  $b>\frac{1}{4}$ no opinion wins.}
\label{main-4}
\end{figure}

The topology of the fixed points as a function of the common density $b$ of both side inflexibles is shown in Fig. (\ref{main-4}). It is rather different from the one sided inflexibles of Fig. (\ref{main-1}). 

\begin{figure}
\centering
\includegraphics[width=.5\textwidth]{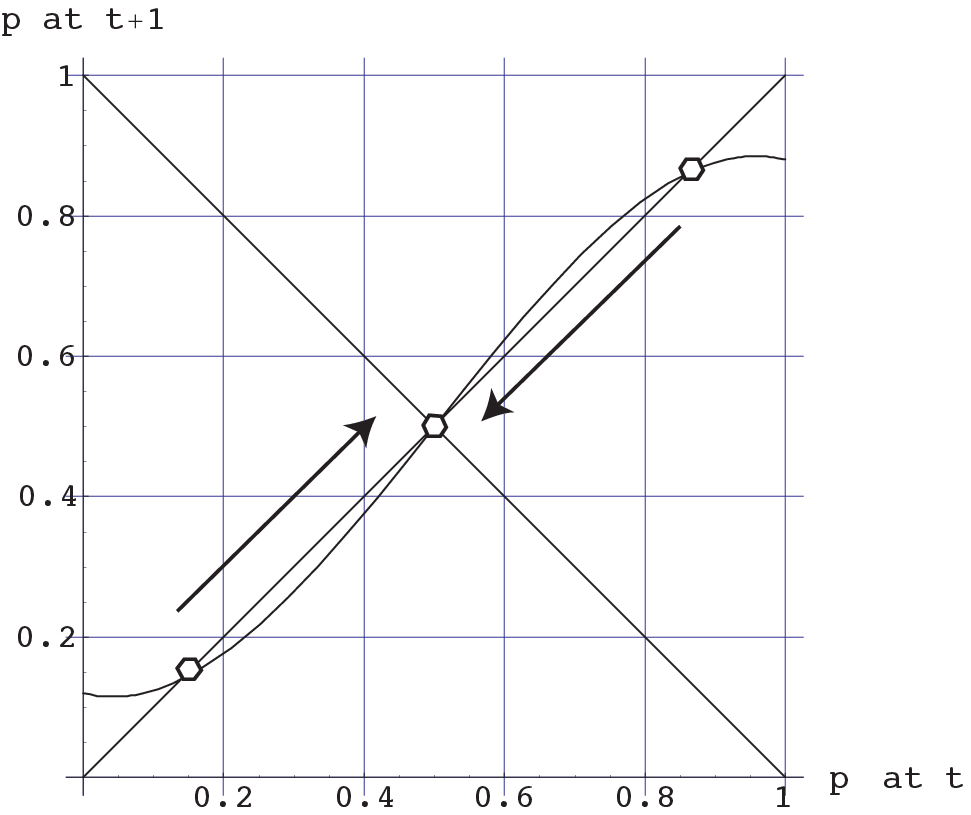}\hfill
\includegraphics[width=.5\textwidth]{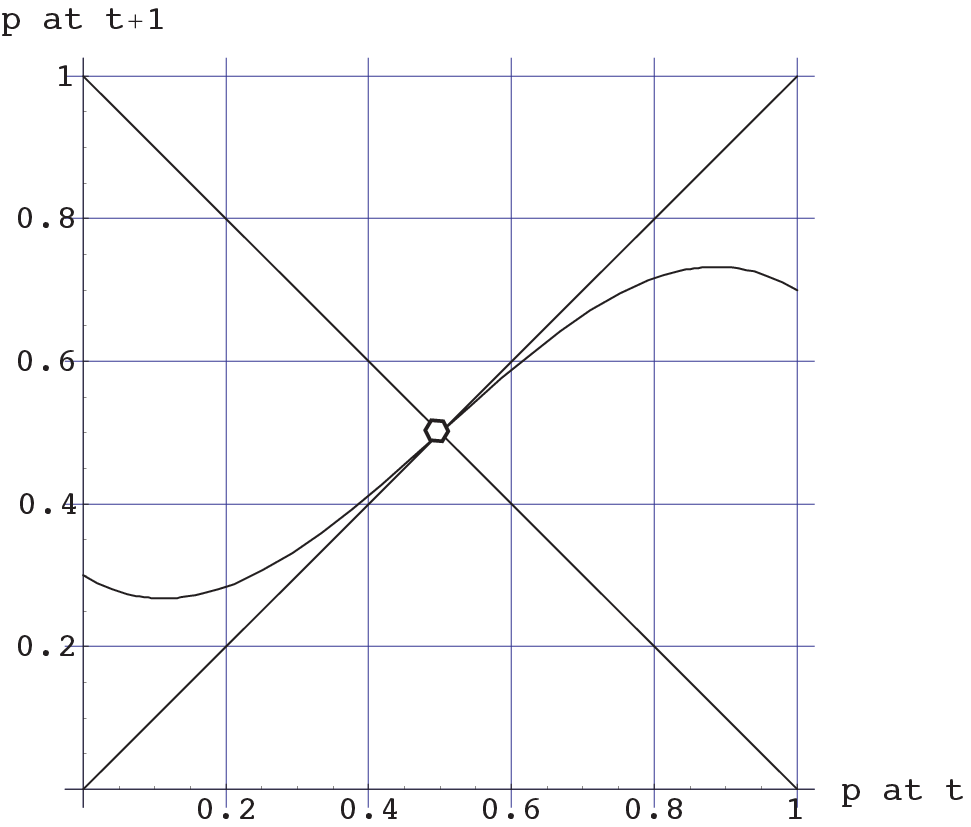}
\caption{One sided inflexibles. The left  part corresponds to $x<0.17$ of inflexibles in favor of opinion A. The two arrows along the diagonal show the directions in which the two attractors move when the equal densities of  inflexibles is increased. The right part shows the case of $x>0.17$, which always yields a stable hung fifty-fifty equilibrium.}
\label{main-5}
\end{figure}

The variation of $p_{t+1}$ as a function of $p_t$ is shown in Fig.  (\ref{main-5}) for the two regimes $b<\frac{1}{4}$ and $b>\frac{1}{4}$. It is worth to notice that the presence of contrarians leads to the same scenario \cite{contra-1}. However, the bare mechanism and its psycho-sociological meaning are quite different. In addition, while $17\%$ of contrarians are necessary to reverse the dynamics, $2\times 25\%=50\%$ of infexibles are needed to accomplish the same reversal. A throrough study of the combined effect of simultaneous contrarians and inflexibles is under investigation \cite{frans}. Nevertheless, it is shown below that this similarity holds only for the case of equal densities of inflexibles for each opinion.

It is certainly realistic to consider inflexibles on both sides, but the symmetric hypothesis is peculiar. To account for the numerous situations, which exhibit different densities of inflexibles, we now study the effect of a discrepancy in $a$ and $b$. 

It is thus the general form of Eq. (\ref{b-x}) which has to be solved to determine its associated fixed points. It yields the cubic Equation 
\begin{equation}
y_t^3+A y_t+B=0 ,
\label{cubic}
\end{equation}
which can be solved analytically with $y_t\equiv p_t-\frac{3+x}{6}$, $A\equiv \frac{1+2b+2x}{2}-\frac{(3+x)^2}{12}$ and $B\equiv -\frac{b+x}{2}+\frac{(3+x)(1+2b+2x)}{12}-\frac{(3+x)^3}{108}$. The solution depends on the sign of the discriminant
\begin{equation}
D=\frac{A^3}{27}+\frac{B^2}{4}.
\label{D}
\end{equation}

Being interested in the nature of the associated dynamics what matters is the number of real roots. Their respective formulations being rather anesthetic formulas in $b$ and $x$, we do not explicit them. But we note that for $D<0$ there exists three distinct real solutions, for $D=0$ there are three real solutions of which at least two are equal, and for $D>0$ there are one single real root and two imaginary roots.

\begin{figure}[t]
\centering
\includegraphics[width=.6\textwidth]{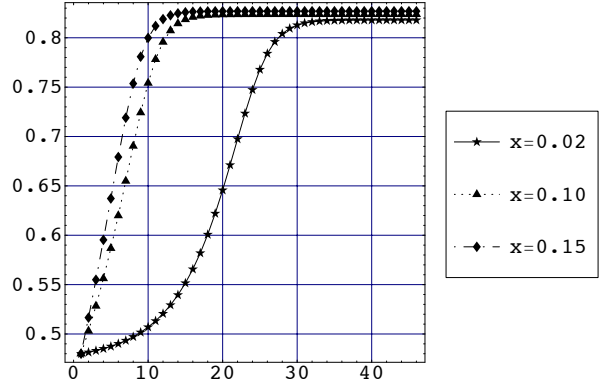}
\caption{Evolution of an initial A support $p_{t}=0.48$ (ordinate) as a function of repeated updates whose number is put on the abscisse. Three different series are shown for respectively $x=0.02, 0.10, 0.15$ with $b=0.15$. The two extreme cases $x=0.02$ and $x=0.15$ yields a similar dynamics. However in the first case an initial $p_{t}=0.46$ would lead the the B victory at contrast with the second case where A wins always.
}
\label{asym-x}
\end{figure}

\begin{description}
\item[(i)] 

The first case of three real solutions ($D<0$) corresponds to the existence of a separator and two attractors as shown in the left part of Fig.  (\ref{asym}). Any positive positive $x$ (more inflexibles in favor of opinion A), shifts the separator below $50\%$ as in the case of one sided inflexible. For instance $b=0.15$ and $x=0.02$ yield $p_{B,a}=0.22$, $p_{c}=0.47$ and $p_{A,b}=0.82$. A $2\%$ difference in inflexible produces a substantial unbalance of  the democratic frame of the public debate since the A opinion needs to start with an initial support larger than $47\%$ to be sure to win an associated election provided the campaign duration is long enough. 

\begin{figure}
\centering
\includegraphics[width=.3\textwidth]{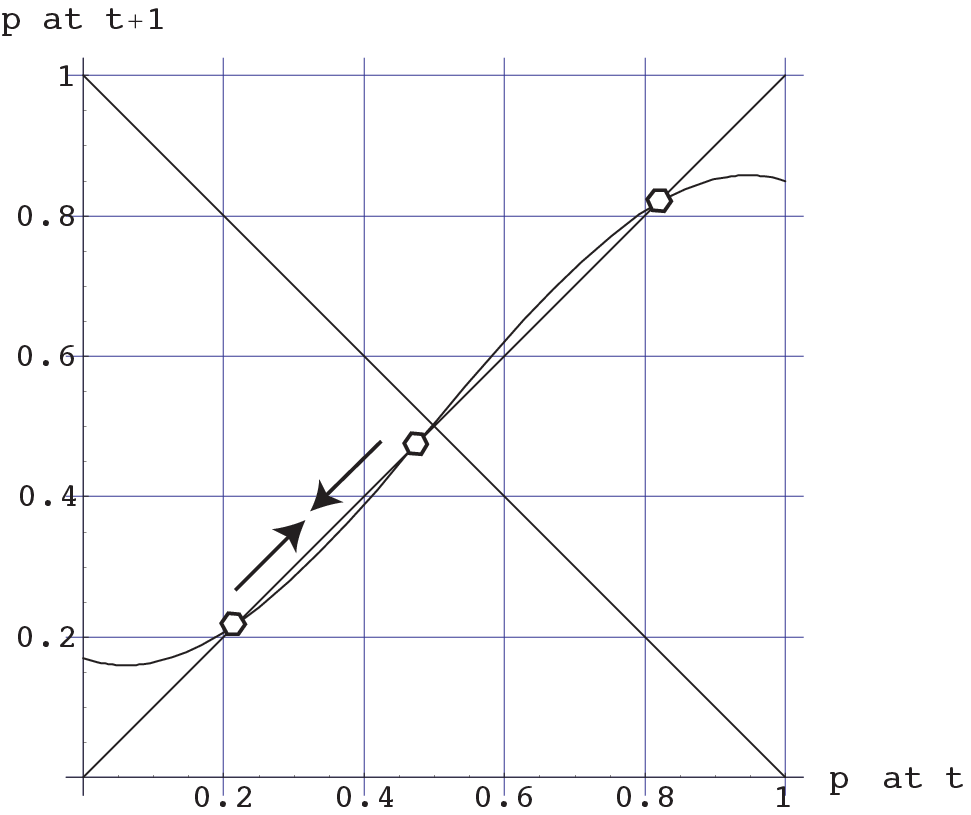}\hfill
\includegraphics[width=.3\textwidth]{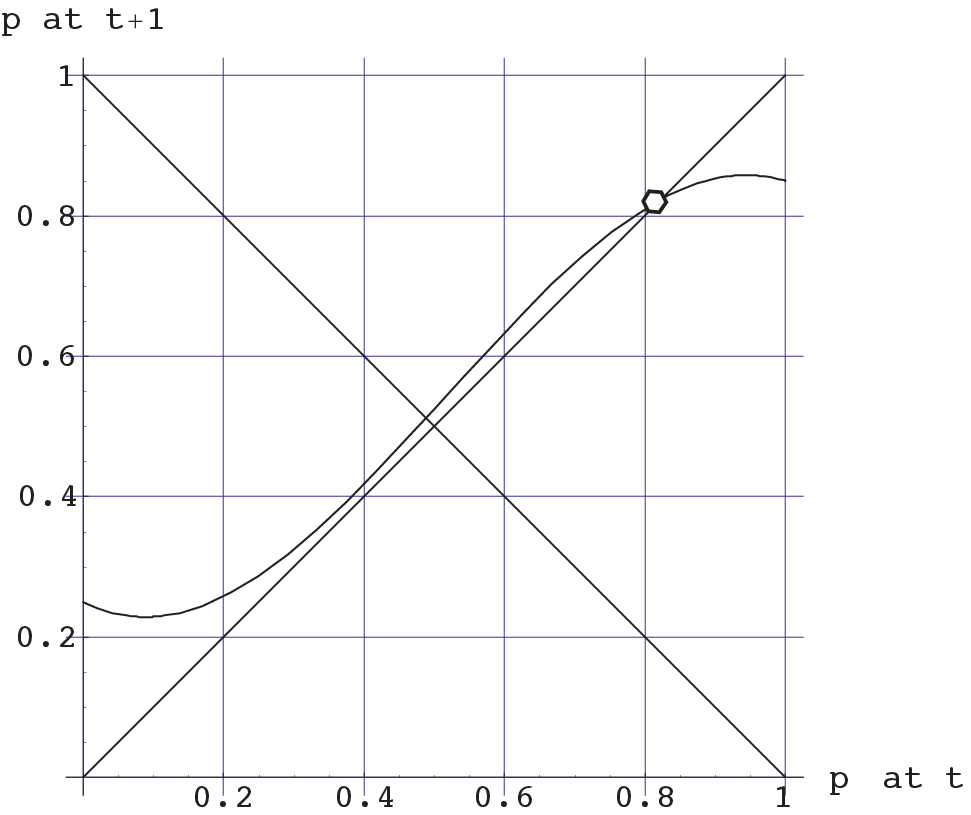}\hfill
\includegraphics[width=.3\textwidth]{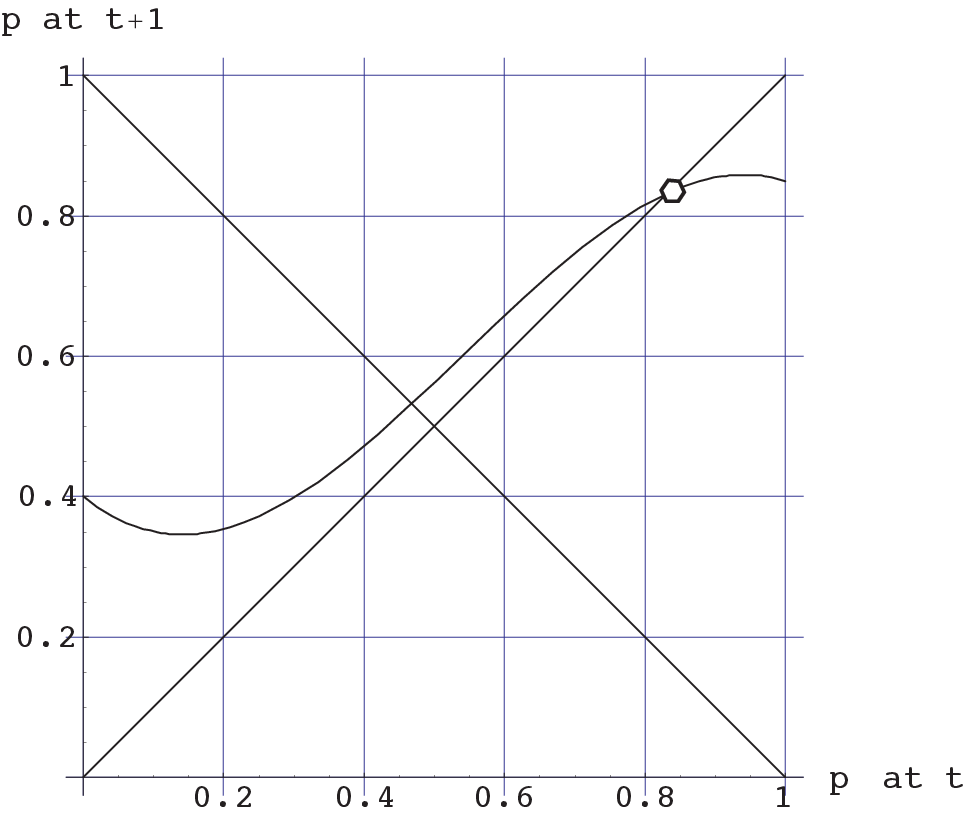}
\caption{Two unequal side inflexibles. The left  part corresponds to $b=0.15, x=0.02$ with three fixed points $p_{B,a}=0.22$ (attractor), $p_{c}=0.47$ (separator) and $p_{A,b}=0.82$ (attractor).   The two arrows along the diagonal show the directions in which the two attractors move when the difference $x$ in densities of  inflexibles is increased. In the middle part $b=0.15$ and $x=0.10>x_c=0.055$ putting the dynamics in the case with the single fixed point $p_{A,b}$ (attractor). The flow is very slow. The right part shows a larger value  $x=0.15$ with still $b=0.15$, which accelerate the converging towards the unique attractor of the dynamics.}
\label{asym}
\end{figure}

For instance, an initial $p_t=0.48$  leads to the series 
$p_{t+1}=0.481, p_{t+2}=0.483, p_{t+3}=0.485, p_{t+4}=0.487, p_{t+5}=0.490, p_{t+6}=0.493, p_{t+7}=0.497$ and  $p_{t+8}=0.502$. Eight updates are necessary to cross the winning bar of fifty percent, i.e. to gain $2.2\%$. To reach a higher score requires more updates with the follow up of $p_{t+9}=0.507, p_{t+10}=0.513, p_{t+11}=0.521, p_{t+12}=0.529, p_{t+13}=0.539, p_{t+14}=0.551, p_{t+15}=0.566, p_{t+16}=0.582, p_{t+17}=0.601$.  Nine additional updates makes the support in favor of A to exceed sixty percent. The majority reversal is here much slower than in the precedent cases. 

It is worth to emphasize that the initial value $p_t=0.46<p_c$ leads to the victory of the B opinion since it starts below the separator located at $p_c=0.47$.  By symmetry, a negative value $x=-0.02$ with the initial value $p_t=0.52$ yields the advantage to opinion B which wins the majority with the same above dynamics.

\item[(ii)]  
Furthermore, given $b$ and increasing $x>0$ results in a continuous shrinking of the distance  between the separator $p_c$ and the mixed phase attractor $p_{B,a}$. At some threshold value $x_c$ both fixed points coalesce. We are then in the second case with two real solutions whose one is double ($D=0$). At reverse, for $x<0$ it is $p_c$ and $p_{A,b}$ which coalesce at $x=-x_c$. Above choice $b=0.15$ yields $x_c=0.055$.

\item[(iii)] 

Afterwards for $x>x_c$ the two fixed points which have coalesced disappear leaving $p_{A,b}$ as the single attractor of the dynamics. To disappear means they became imaginary, we are in the third case $D>0$ with one single real solution $p_{A,b}$.

For $x>x_c$, in the vicinity of $x_c$ the flow is very low as seen in the middle part of Fig.  (\ref{asym}) where we have the set ($b=0.15, x=0.10$). The dynamics in the third case with only one unique fixed point, an attractor and above initial value  $p_t=0.48$  yields now the series $p_{t+1}=0.503, p_{t+2}=0.528, p_{t+3}=0.556, p_{t+4}=0.587, p_{t+5}=0.620$. 

One single update is now sufficient to rise the minority opinion A to the status of majority as compared to eight updates above. Only four additional updates reach the  sixty percent bar instead of the previous nine. The majority reversal has been accelerated. 

Going to the set ($b=0.15, x=0.15$)  makes the dynamics faster as exhibited in the right part of Fig.  (\ref{asym}). We now have from $p_t=0.48$  the series $p_{t+1}=0.517, p_{t+2}=0.555, p_{t+3}=0.595, p_{t+4}=0.637, p_{t+5}=0.679$. 

As soon as $\pm x_c$ are reached the dynamics  ineluctably leads the opinion which have the surplus of inflexibles to invade the majority of the population (A for $x_c$ and B when  $\pm x_c$ ). The above three different series for $b=0.15$ and $x=0.02, 0.10, 0.15$  are reproduced in Fig.  (\ref{asym-x}).

\end{description}

It thus appear to be of a central importance to determine the value of $x_c$ given the value of $b$. Once the associated opinion reached a surplus of inflexibles $x_c$ it eventually wins the election with certainty. To achieve this goal, we need to solve the Equation $D=0$ as a function of the variable $x$, $b$ being a fixed parameter, where $D$ is given by Eq. (\ref{D}). 

\begin{figure}[t]
\centering
\includegraphics[width=.6\textwidth]{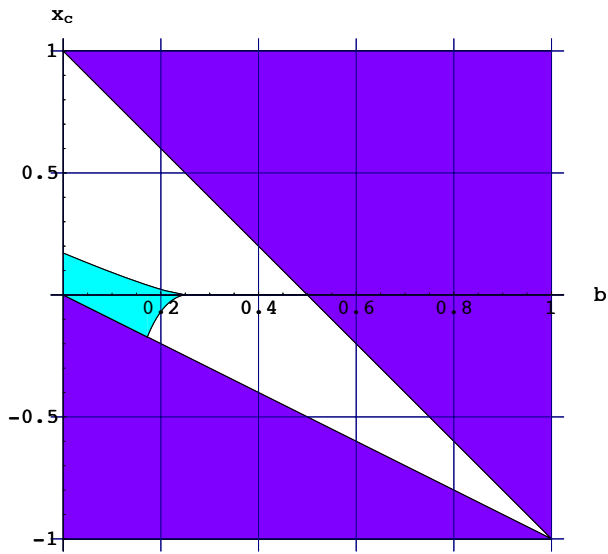}
\caption{The dynamics map. The white triangle delimited by $0\leq b\leq 1$ and $-b\leq x\leq 1-2b$ shows the accessible range for the respective values of $b$ and $x$. Within the accessible area, the left grey area corresponds to region where $D<0$, and the dynamics is monitored by a separator and two attractors with $x_{c2}<x<x_{c1}$ where $x_{c2}\leq 0$ and $x_{c}\geq 0$. Outside this closed area, the dynamics is driven by a  single attractor. }
\label{xc}
\end{figure}

Performing a Taylor expansion of Eq. (\ref{D}) in power of $x$ at order 2 leads to the solutions
\begin{equation}
x_{c1,c2}=\frac{3 - 24b + 48b^2 \mp 2(-1+ 4b)^{3/2}\sqrt{-2+ b +b ^2}}{1 - 32b + 4b^2},
\label{xc-2}
\end{equation}
which are shown in Fig  (\ref{xc}) together with the available values for $(b,x)$ constrained by the frontiers $0\leq b\leq 1$ and $-b\leq x\leq 1-2b$. The positive value $x_{c1}$ exists the range $0\leq b \leq \frac{1}{4}$ while for the negative value $x_{c2}$ it is the range $3-2\sqrt{2} \approx 0.17 \leq b \leq \frac{1}{4}$. 

In the region $x_{c2}<x<x_{c1}$, $D<0$ which yields a separator and two attractors. At odd, outside this closed area and with $-b <x<1-2b$, we have  $D<0$ with one single attractor. The case $x>0$ guarantees the A victory while $x<0$ grants the B victory. The various domains are shown in Fig  (\ref{xc}). It appears that $D>0$ for $b>\frac{1}{4}$. A positive $x$ yields a A victory while a negative $x$ a B victory. The three fixed points coalesce at the unique set $b=\frac{1}{4}, x=0$.

We have singled out the effect of inflexible choices on the democratic opinion forming. An inflexible being an agent who always sticks to its opinion without any shift. At low and equal densities, they prevent the trend towards a total polarization of floaters along one unique opinion. The opinion dynamics is found to lead to a mixed phase attractor with a clear cut majority-minority splitting. Below $25\%$ of equal density  inflexibles for both opinions, the initial majority opinion wins the public debate. At contrast,  beyond $25\%$  the dynamics is reversed and converge towards a fifty-fifty attractor. Therefore an equal density of inflexibles produces effects which can also be achieved by sufficiently low densities of contrarians \cite{contra-1}.

However, even a very small asymmetry in the respective inflexibles densities upsets the balanced character of above results. At a very low difference, the main effect is to shift the separator from fifty percent to a lower value at the advantage of the larger inflexible opinion. It also increases its incompressible minority support. Moreover, an excess in inflexibles beyond some small threshold $x_c$, which depends on $b$, grants the victory to the beneficiary opinion. In this regime there exists only one single attractor, which drives the corresponding  opinion to an overwhelming majority. Nevertheless it is worth to emphasize that the associated dynamics may become  rather slow.

Fig  (\ref{xc}) sums up our results. it allows to determine which strategy is best for a given opinion to win the public debate competition.  It appears that the decisive  goal should be to get a lead, even small, in the respective inflexible densities. It immediately  produces the substantial advantage to lower the separator from $50\%$. A larger difference in inflexibles, whose amplitude varies as a function of the other opinion support, guarantees the winning of the campaign, and eventually the follow up election.

On this basis we plan to extend our study to larger size update groups. We also plan to combine both effects of contrarians and inflexibles to study the dynamics of floaters \cite{frans}.


 \end{document}